# Exploration of the computational model and the focusing process with a Flat Multi-channel Plate and a Curved Multi-channel Plate in the MATLAB

Mo Zhou[1,2,3], Kai Pan[1,2,3,6], Tian-Cheng Yi[4,5], Xing-Fen Jiang[4,5], Bin Zhou[4,5], Jian-Rong Zhou[4,5], Xue-Peng Sun[1,3], Song-Ling Wang[4,5], Bo-Wen Jiang[6], Tian-Xi Sun[1,2,3] and Zhi-Guo Liu[1,2,3]

[1]*College of Nuclear Science and Technology, Beijing Normal University, Beijing 100875, China*
[2]*Beijing Key Laboratory of Applied Optics, Beijing 100875, China*
[3]*Key Laboratory of Beam Technology, Ministry of Education, Beijing 100875, China*
[4]*Spallation Neutron Source Science Center, Dongguan 523803, China*
[5]*Institute of High Energy Physics, Chinese Academy of Sciences, Beijing 100049, China*
[6]*North Night Vision Technology (Nanjing) Research Institute Co.,Ltd, Nanjing 211106,China*

**Abstract：**
By simulating the X-ray paths and the Chapman Model of a flat multi-channel plate and a curved multi-channel plate in the MATLAB, the field of view, local reflection efficiency, spherical aberration, point-spread function, collection efficiency of incident X-ray and peak-to-background ratio on the focal plane of the two devices were compared. At the same time, the advantages and disadvantages of the flat multi-channel plate and the curved multi-channel plate were compared.

1. **Introduction：**
   Similar to the traditional X-ray polycapillary lens, the square polycapillary slice lens (SPSL) can control the X-ray beam based on the principle of total reflection of X-rays on the multi-channel plate internal surface.[1-3] Compared with diffractive X-ray control lenses, total reflection X-ray control lenses are more suitable for broadband beams.[6-8] Basis of this advantage, the SPSL can realize the X-ray focusing and imaging functions of a large field of view (FOV), so it was widely used in the field of astronomical observation.[4-5] A single SPSL was composed of micron-scale square channels periodically arranged on the order of millions. Mathematical modeling is the premise of analyzing its X-ray regulation capability. At present, the X-ray tracing method is mainly used for the design of such optical control lenses.[9-12]

   In this paper, a new type of curved multi-channel plate was proposed. First, a new simulation method was proposed to optimize the interface reflectivity of the curved SPSL. It was mathematically modeled and its optical behavior was analyzed. Secondly, the influence of the parameters of a flat multi-channel plate (FMCP) and a curved multi-channel plate (CMCP) on the lenses design were compared and analyzed.

**2. Two-Parameter Reflectivity Curve Model:**

Regarding the computational model of the SPSL, the most classic is the "Chapman Model" proposed by H.N.Chapman et al. of the Melbourne University. In the optical structure design of the SPSL, we followed this model and made related improvements. The functional simplification of the interface reflectance curve in the "Chapman" model is related to only one parameter of the incident ray energy. In order to make the reflectance model more in line with the actual situation and better reflect the optical performance of the curved lens, it is necessary to modify the interface reflectance model. In the range of hard X-ray energy, with the increase of energy, the change of reflectance with incident angle can be described in the form of a step function, which is more in line with the actual situation; in the range of soft X-ray, the change of reflectance with incident angle can be described in the form of a linear function. The change of angle, when the grazing incidence angle is small, the simplified interface reflectance curve is in better agreement with the actual curve. At the same time, the reflectivity of the interface in the model is affected by the random roughness and will drop significantly. The revised interface reflectance model is mainly related to the refractive index and surface roughness of the material, and is also called two-parameter reflectivity curve model (TRCM). The specific elaboration of this model of interface reflectivity is as follows:

For the FMCP, the two factors in TRCM that reflect the refractive index and surface roughness of the material are the critical angle of total reflection $\gamma_c$ and the average reflectance $\bar{R}$, respectively. $\gamma_c$ can be calculated by:

$$\gamma_c = \lambda \sqrt{\frac{r_e}{2\pi} N_{av} \rho_m K}$$

$$K = 2 \frac{\sum_i c_i \left(f_{0i} + f_i^{'}(\lambda)\right)}{\sum_i c_i M_i} \quad (2\text{-}1)$$

Where $\lambda$ is the X-ray wavelength, $r_e$ is the classical electron radius, $N_{av}$ is the Avogadro constant, $\rho_m$ is the mass density of the material, $c_i$ is the mass percentage of the $i$ element in the composite, $f_{0i} + f_i^{'}(\lambda)$ is the atomic scattering factor and $M_i$ is the atomic mass of the $i$ element.

The interface reflectance $R(\gamma, \lambda)$ of X-ray of a given wavelength can be obtained by the Fresnel formula. $R(\gamma,\lambda)$ will be integrated over the critical angle to obtain the average interface reflectance $\overline{R}(\lambda)$ over the critical angle of total reflection:

$$\overline{R}(\lambda)=\frac{1}{\gamma_c(\lambda)}\int_0^{\gamma_c(\lambda)} R(\theta_i,\lambda)d\theta_i \qquad (2\text{-}2)$$

Considering the influence of random roughness on the reflectivity, the Debye-Waller factor is introduced, and $\overline{R}(\lambda)$ is integrated within the critical angle range, and the simplified reflectivity function model under the two-factor model can be obtained:

$$\Re(\theta_i,\lambda)=\begin{cases}1+2(\overline{\Re}(\lambda)-1)\theta_i/\gamma_c(\lambda), & \theta_i \leq \gamma_c(\lambda)\\ 0, & \theta_i > \gamma_c(\lambda)\end{cases} \qquad (2\text{-}3)$$

When the material's absorption of X-rays is low and the surface roughness is small, formula is consistent with the simplified function of hard X-ray reflectance in the Chapman model; when the material's absorption coefficient increases, its Consistent with a linear reduction function for soft X-rays. In particular, near the absorption edge of the material, due to the sudden change of the absorption coefficient and the combined action of the surface roughness, the average reflectance $\overline{R}(\lambda)$ will be less than 0.5, and $\Re(\theta_i,\lambda)$ is a negative number at this time. In order to make the reflectivity positive, it is necessary to define when $\overline{R}(\lambda)<0.5$:

Figure 1(a) is the curve of the interface reflectance of Ir with a completely smooth surface as a function of the incident X-ray energy when the incident ray energy is 6.4 keV. The black line is the reflectance curve calculated by the Fresnel formula, and the red line is the simplified curve calculated using the two-factor model. In Figure 1(b), the black line is the change of the average interface reflectance with incident energy without roughness, and the red line is the change of the average interface reflectance with the incident energy when the roughness root mean square is 2 nm. With the increase of X-ray energy, the influence of interface roughness on the average interface reflectivity increases.

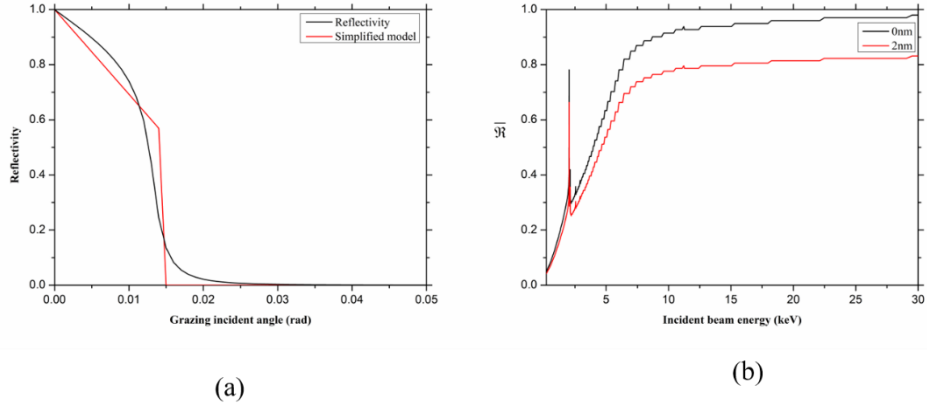

Figure 1. (a) Simplified curve of interface reflectance of Ir under two-factor model; (b) Effect of different roughness on average interface reflectance

From the above, the simplified reflectivity curve using the two-factor model can well reflect the reflection of grazing incident rays on surfaces with different roughness such as Au, Ir, etc. with high atomic number. Compared with the simplified functional form of energy segments in the "Chapman" model, the two-factor model is more uniform and accurate.

## 3. X-ray Tracing Simulation
### A. Lenses Basic Geometry

When constructing the basic geometric structure of the curved lens, in the lateral direction, the lens is divided into four symmetrical regions according to quadrants. During the calculation, only the first quadrant is considered, and the results of other quadrants are reversed according to the symmetry. The multi-channels on the axis are counted separately. The multi-channels located at the geometric center of the lens are numbered index $(0,0)$, located in the i layer, and the multi-channels in the j column are numbered $(i,j)$, and the overall lens has n layers and n columns of multi-channels. Figure 2(a) shows the basic geometry of the lens when the multi-channel is square and the wall thickness is considered. In the one-dimensional direction, the included angle between the central axes of two adjacent channels is $\alpha_{sc} = (d_{sc} + 2d_{sc\_wall})/R$. For the channel numbered $(i,j)$, in the two-dimensional direction, record the angle between its central axis and the main axis Z as $(\phi_{sc\_i}, \phi_{sc\_j})$, there are $\phi_{sc\_i} = i\alpha_{sc\_i}$ and $\phi_{sc\_j} = j\alpha_{sc\_j}$. The field angles of the lens in the horizontal and vertical directions are $(n+1)\alpha_{sc\_i}$ and $(n+1)\alpha_{sc\_j}$, respectively. Figure 2(b) shows the basic geometry of the lens when the multi-channel is tapered and the wall thickness is considered. Since both the channel wall and the hollow part are conical, in one-dimensional direction, the angle between the central axes of two adjacent channels is the sum of the cone apex angles corresponding to the two parts and the center of curvature, namely $\alpha_{tc} = \alpha_{tc\_hollow} + 2\alpha_{tc\_wall}$. The field of view angles of the lens in the horizontal and vertical directions are $(n+1)\alpha_{tc\_i}$ and $(n+1)\alpha_{tc\_j}$, respectively.

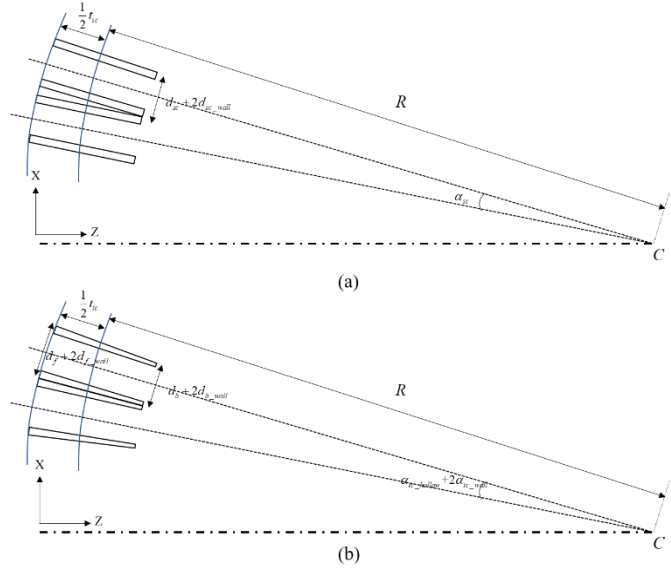

Figure 2. Basic geometry of curved lens in one dimension: (a) the channel is square; (b) the channel is conical

The coordinate schematic diagram of 2(b) is simplified, and the simplified diagram is Figure 3. In Figure 3, the distance between the coordinate origin and the center of curvature of the lens is $l_{oc} = (R+\tfrac{1}{2}t)\cos((n+1)\alpha_i)$. When the multi-channel is square, for a single multi-channel in the first quadrant $(i,j)$, the space equations of the upper, lower, left and right reflective inner surfaces are:

$$\begin{cases} up: -\tan(j\alpha_{sc\_j})(z-l_{oc}) + \dfrac{d_{sc}}{2\cos(j\alpha_{sc\_j})} = y \\[4pt] down: -\tan(j\alpha_{sc\_j})(z-l_{oc}) - \dfrac{d_{sc}}{2\cos(j\alpha_{sc\_j})} = y \\[4pt] left: -\tan(i\alpha_{sc\_i})(z-l_{oc}) + \dfrac{d_{sc}}{2\cos(i\alpha_{sc\_i})} = x \\[4pt] right: -\tan(i\alpha_{sc\_i})(z-l_{oc}) - \dfrac{d_{sc}}{2\cos(i\alpha_{sc\_i})} = x \end{cases} \quad (3\text{-}1)$$

When the multi-channel is tapered, for a single multi-channel in the first quadrant $(i,j)$, the spatial equations of the upper, lower, left and right reflective inner surfaces are:

$$\begin{cases} up: -\tan\left(\left(j+\dfrac{1}{2}\right)\alpha_{tc\_j}\right)(z-l_{oc}) = y \\ down: -\tan\left(\left(j-\dfrac{1}{2}\right)\alpha_{tc\_j}\right)(z-l_{oc}) = y \\ left: -\tan\left(\left(i+\dfrac{1}{2}\right)\alpha_{tc\_i}\right)(z-l_{oc}) = x \\ right: -\tan\left(\left(i-\dfrac{1}{2}\right)\alpha_{tc\_i}\right)(z-l_{oc}) = x \end{cases} \quad (3\text{-}2)$$

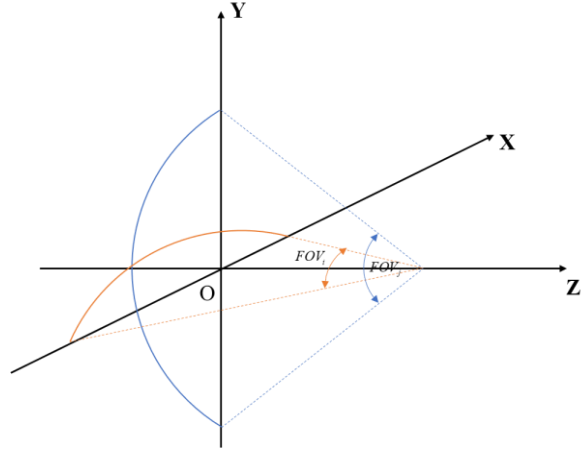

Figure 3. Schematic diagram of simplified coordinates of curved lens

Due to manufacturing errors, curved lenses actually have many structural defects. The mathematical description for the four inner walls of the tapered multi-channel lens is:

$$\begin{cases} up: x(\tan\theta_{adlu}\sin\theta_l + \cos\theta_l)\sin\theta_t + y(\tan\theta_{adlu}\sin\theta_l + \cos\theta_l)\cos\theta_t \\ \quad + (z-l_{oc})(\tan\theta_{adlu}\cos\theta_l - \sin\theta_l) = 0 \\ down: x(\tan\theta_{adld}\sin\theta_l + \cos\theta_l)\sin\theta_t + y(\tan\theta_{adld}\sin\theta_l + \cos\theta_l)\cos\theta_t \\ \quad + (z-l_{oc})(\tan\theta_{adld}\cos\theta_l - \sin\theta_l) = 0 \\ left: x(\cos\theta_t - \tan\theta_{adtl}\sin\theta_l\sin\theta_t) + y(\sin\theta_t + \tan\theta_{adtl}\sin\theta_l\cos\theta_t) \\ \quad + (z-l_{oc})(\tan\theta_{adll}\cos\theta_l) = 0 \\ right: x(\cos\theta_t - \tan\theta_{adtr}\sin\theta_l\sin\theta_t) + y(\sin\theta_t + \tan\theta_{adtr}\sin\theta_l\cos\theta_t) \\ \quad + (z-l_{oc})(\tan\theta_{adlr}\cos\theta_l) = 0 \\ \theta_{adlu} = \left(j_{axis\_de} + \dfrac{1}{2}\right)\alpha_{tc\_j}, \theta_{adld} = \left(j_{axis\_de} - \dfrac{1}{2}\right)\alpha_{tc\_j} \\ \theta_{adtl} = \left(i_{axis\_de} + \dfrac{1}{2}\right)\alpha_{tc\_i}, \theta_{adtr} = \left(i_{axis\_de} - \dfrac{1}{2}\right)\alpha_{tc\_i} \end{cases} \quad (3\text{-}3)$$

The lateral axis deviation angle is recorded as $\theta_t$, the axial axis deviation angle is $\theta_l$, the upward inclination angle is positive, and the downward angle is negative. Assume that the starting number of the multi-channel with two-dimensional axis deviation is $(i_{axis\_de}, j_{axis\_de})$

The starting number of the multi-channel with displacement defects is $(i_{disp}, j_{disp})$, and the lateral displacements in the X and Y axis directions are $\Delta d_x(i_{disp}, j_{disp})$ and $\Delta d_y(i_{disp}, j_{disp})$, respectively, then the mathematical description of the four walls of the multi-channel with displacement defects is:

$$\begin{cases} up: -\tan\left(\left(j+\frac{1}{2}\right)\alpha_{tc\_j}\right)(z-l_{oc}) + \Delta d_y(i_{disp}, j_{disp}) = y \\ down: -\tan\left(\left(j-\frac{1}{2}\right)\alpha_{tc\_j}\right)(z-l_{oc}) + \Delta d_y(i_{disp}, j_{disp}) = y \\ left: -\tan\left(\left(i+\frac{1}{2}\right)\alpha_{tc\_i}\right)(z-l_{oc}) + \Delta d_x(i_{disp}, j_{disp}) = x \\ right: -\tan\left(\left(i-\frac{1}{2}\right)\alpha_{tc\_i}\right)(z-l_{oc}) + \Delta d_x(i_{disp}, j_{disp}) = x \end{cases} \quad (3\text{-}4)$$

Extrusion defect, the normal vector of the plane equation of the multi-channel with extrusion defect is rotated at an angle of $\theta_{extr}$, and the channel number is $(i_{extr}, j_{extr})$. The mathematical description of the four walls of the multi-channel is:

$$\begin{cases} up: -\tan\left(\left(j+\frac{1}{2}\right)\alpha_{tc\_j}\right)(z-l_{oc}) = y\sin\theta_{extr} + x\cos\theta_{extr} \\ down: -\tan\left(\left(j-\frac{1}{2}\right)\alpha_{tc\_j}\right)(z-l_{oc}) = y\sin\theta_{extr} + x\cos\theta_{extr} \\ left: -\tan\left(\left(i+\frac{1}{2}\right)\alpha_{tc\_i}\right)(z-l_{oc}) = x\sin\theta_{extr} + y\cos\theta_{extr} \\ right: -\tan\left(\left(i-\frac{1}{2}\right)\alpha_{tc\_i}\right)(z-l_{oc}) = x\sin\theta_{extr} + y\cos\theta_{extr} \end{cases} \quad (3\text{-}5)$$

In addition to the above three common manufacturing errors, curved lenses also have curvature deviation and surface deviation. As shown in Figure 4 (a), after the light emitted by the point light source passes through the lens, due to the actual curvature deviation of the lens, it finally converges before or after the ideal focal plane. Lens curvature deviation is common in curved lenses with square multi-channels. Figure 4(b) shows the focus error caused by the detector surface. Since the curved lens has a wide field of view, it can converge parallel light sources in different directions, and the final ideal imaging plane is a spherical surface. In fact, the surface type of the current general detector is a flat plate type. When the parallel light source is incident in the direction of the black beam as shown in the figure, the flat-panel focal spot has little effect; when the parallel light source is incident in the direction of the yellow beam in the figure, there is a declination from the Z-axis, on the flat-panel detector, the focal spot The plaques will appear asymmetrical and severely diffuse. For the simulation of curvature deviation, it is only necessary to move the coordinates of the detection surface away from the ideal focal plane in the axial direction; for the deviation caused by the detector surface shape, the light source needs to be deviated from the Z axis by an angle. In addition, in the actual modeling process, since the focal plane of the curved lens is spherical, the focal spot information needs to be reflected by the projection method from the spherical surface to the plane.

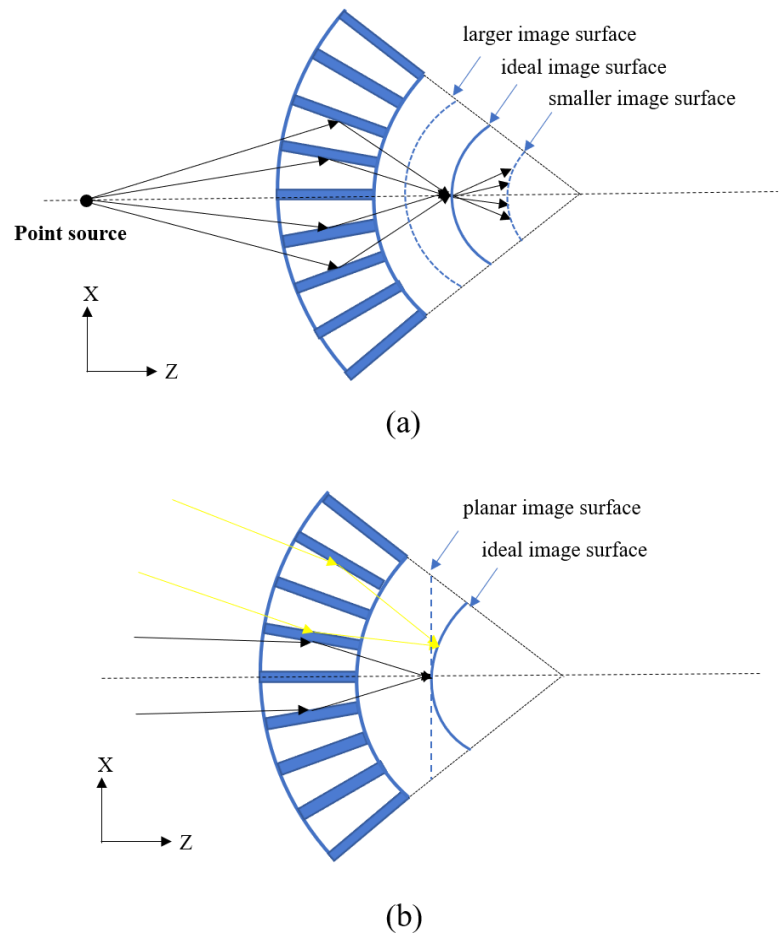

Figure 4. Curved lens (a) curvature deviation; (b) detector surface deviation

**B. X-ray tracing process inside the lenses**

Taking a single point light source $p_s(x_s, y_s, z_s)$ as an example, the X-ray tracing process in a curved lens is described. Let the energy of the light emitted by the light source be $E_s$, and the initial intensity count of a single X-ray is 1. For a surface light source, if its light intensity has a distribution, the initial light intensity of the light is proportional to its distribution function. Taking the Gaussian distribution as an example, the initial intensity of a single X-ray is related to the lateral coordinate of the light source, which is $I_o(x_s, y_s)$. For a single multi-channel $(i, j)$, let its x-axis and y-axis coordinate ranges be $[x_{pore\_left}, x_{pore\_right}]$ and $[y_{pore\_left}, y_{pore\_right}]$, respectively. If the multi-channel shape is square:

$$\begin{cases} x_{pore\_left} = \left(R + \frac{1}{2}t_{sc}\right)\sin(i\alpha_{tc\_i}) + \frac{d_{sc}}{2\cos(i\alpha_{tc\_i})} \\ x_{pore\_right} = \left(R + \frac{1}{2}t_{sc}\right)\sin(i\alpha_{tc\_i}) - \frac{d_{sc}}{2\cos(i\alpha_{tc\_i})} \\ y_{pore\_left} = \left(R + \frac{1}{2}t_{sc}\right)\sin(j\alpha_{tc\_j}) + \frac{d_{sc}}{2\cos(j\alpha_{tc\_j})} \\ y_{pore\_right} = \left(R + \frac{1}{2}t_{sc}\right)\sin(j\alpha_{tc\_j}) - \frac{d_{sc}}{2\cos(j\alpha_{tc\_j})} \end{cases} \quad (3\text{-}6)$$

If the multi-channel is tapered, there are:

$$\begin{cases} x_{pore\_left} = \left(R + \frac{1}{2}t_{sc}\right)\sin\left(\left(i+\frac{1}{2}\right)\alpha_{tc\_i}\right) \\ x_{pore\_right} = \left(R + \frac{1}{2}t_{sc}\right)\sin\left(\left(i-\frac{1}{2}\right)\alpha_{tc\_i}\right) \\ y_{pore\_left} = \left(R + \frac{1}{2}t_{sc}\right)\sin\left(\left(j+\frac{1}{2}\right)\alpha_{tc\_j}\right) \\ y_{pore\_right} = \left(R + \frac{1}{2}t_{sc}\right)\sin\left(\left(j-\frac{1}{2}\right)\alpha_{tc\_j}\right) \end{cases} \quad (3\text{-}7)$$

Randomly select the entry coordinate $(x_{pore}(i,j), y_{pore}(i,j), z_{pore}(i,j))$ of the X-ray within the range of $[x_{pore\_left}, x_{pore\_right}]$ and $[y_{pore\_left}, y_{pore\_right}]$. The direction vector $\vec{u_i}$ of the incident X-ray can be determined from the coordinates of the light source and the coordinates of the random entrance.

Considering that the multi-channel has no structural defects, the coordinates of the collision point for the first collision between the incident X-ray and the two types of multi-channels are calculated respectively. The coordinate of a collision point is $(x_{1c}, y_{1c}, z_{1c})$, since the incident X-ray can only collide with one of the four walls. The collision wall can be judged by the constraints:

$$R-\frac{t}{2} \leq \sqrt{x_{1c}^2 + y_{1c}^2 + (z_{1c}-l_{oc})^2} \leq R+\frac{t}{2} \quad (3\text{-}8)$$

Consider the case where the detector surface is spherical. Let the horizontal and vertical fields of view of the spherical detector be $\theta_h$ and $\theta_v$ respectively, the radius of curvature of the detector is $R_d$, and the coordinate of the center of curvature is $(0,0,z_{det})$. The solid angle corresponding to a single detection unit in the two-dimensional direction is $\theta_{pix} \times \theta_{pix} \, (mrad)$, and the entire spherical detector contains $(2N_{det}+1)\times(2N_{det}+1)$ pixel units. Let the central pixel of the detection sphere be numbered $(0,0)$, and for the light detected by the detection unit $(m_{det}, n_{det})$, the horizontal and vertical angles with the Z axis are recorded as $(\theta_{hdet}, \theta_{vdet})$. For a X-ray that satisfies the constraints, its reflection angle after one reflection can be calculated. The reflected light is used as the incident light of the next reflection, and the above process is repeated until the exit or greater than the critical angle of total reflection. By finding the intersection of the last reflected X-ray and the detection surface, the unit number (the number can be negative) of the final X-ray detected by the spherical detector can be obtained:

$$\begin{aligned} m_{det} &= floor\left(\frac{x_f}{R\theta_{pix}}\right), \\ n_{det} &= floor\left(\frac{y_f}{R\theta_{pix}}\right) \end{aligned} \quad (3\text{-}9)$$

A single X-ray is reflected multiple times and transformed using the Mercator projection. The formula for the Mercator orthographic projection is::

$$\begin{cases} X = K\ln\left[\tan\left(\frac{\pi}{4}+\frac{B}{2}\right)\left(\frac{1-e\sin B}{1+e\sin B}\right)^{\frac{e}{2}}\right] \\ Y = K(L-L_0) \\ K = \frac{a^2/b}{\sqrt{1+e^2\cos^2(B_0)}}\cos(B_0) \end{cases} \quad (3\text{-}10)$$

Among them, $a$ is the semi-major axis of the ellipsoid, $b$ is the semi-minor axis of the ellipsoid, $e$ is the eccentricity, B is the dimension, L is the longitude, $B_0$ is the standard dimension $L_0$ and is the standard longitude.

For spherical detectors, its standard latitude and standard longitude are both 0. When performing projection transformation, set its spherical center coordinate as $(0,0,R_{det}/2)$, then its standard longitude and latitude is $(0,0)$, and the corresponding Cartesian coordinate is $(0,0)$. Then the Mercator projection transformation suitable for spherical detectors can be simplified as:

$$\begin{cases} X = RL \\ Y = R\ln\left(\tan\left(\frac{\pi}{4}+\frac{B}{2}\right)\right) \end{cases} \quad (3\text{-}11)$$

Above, by replacing the multi-channel equation without manufacturing defects with an equation that considers the influence of manufacturing defects, the trace simulation can be performed for the situations where there are four kinds of manufacturing errors. The presence of curvature deviations can be simulated by changing the axial coordinates of the detection surface. The deviation caused by the detector surface can be simulated by replacing the spherical detection surface with a flat surface.

## 4. Lenses Performance Parameters
### A. FOV

Since the curved square multi-channel slice lens has a spherically symmetric structure as a whole, a single multi-channel has independent symmetrical optical axes, and these independent optical axes all point to the center of curvature of the lens, which determines that the curved lens has the advantage of a wide field of view. For long-distance targets, splicing multiple curved lenses can realize the collection of light sources in different directions, and finally converge on the spherical focal plane, as shown in Figure 5(a). For a single curved lens, the FOV size is an important optical characterization parameter. The field of view of the curved lens can be divided into a horizontal field of view and a vertical field of view. As shown in Figure 5(b), the effective field of view of the curved lens is mainly related to the number of channels M of the lens and the size of the cone apex $\alpha_{tc/sc}$ of a single channel:

$$\theta_{fov\_eff} = M\alpha_{tc/sc} \quad (4\text{-}1)$$

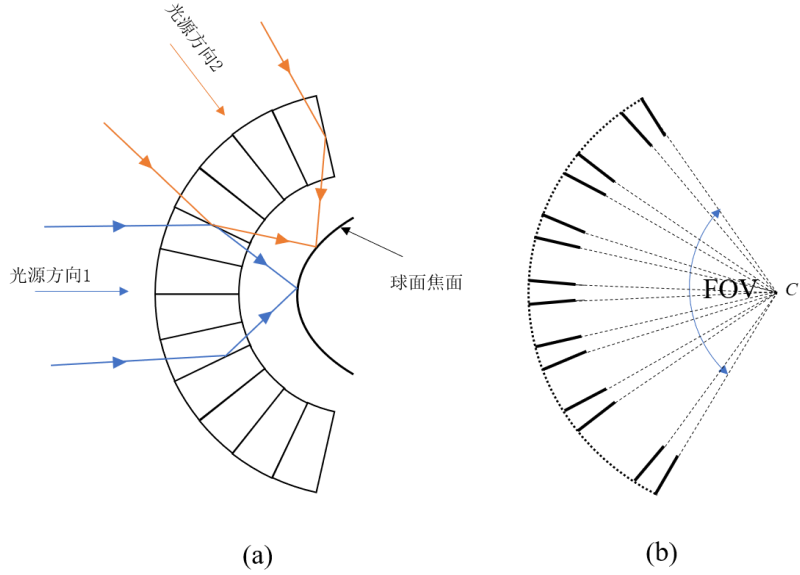

Figure 5. (a) The focusing of parallel light sources in different directions by the curved lens; (b) the field of view of the curved lens intrinsic reflection efficiency

## B. Local Reflection Efficiency

The intrinsic reflection efficiency of the curved square multi-channel lens needs to introduce the curvature factor $\chi$, then the formula becomes:

$$\varepsilon_{n_x,n_y}(\theta_x,\theta_y) = \frac{1}{\bar{d}^2}\prod_{n=1}^{nx} R(\theta_x \chi)\prod_{n=1}^{ny} R(\theta_y \chi)\delta_{n_x}(\theta_x)\delta_{n_y}(\theta_y)$$

$$\bar{d} = \begin{cases} d, & \text{square channel} \\ (d_f + d_b)/2, & \text{tapered channel} \end{cases}$$

(4-2)

Figure 6(a) shows the two-dimensional distribution of the intrinsic efficiency of the tapered multi-channel lens when the incident energy is 1.00 keV. The intrinsic efficiency of the curved lens is symmetric and reaches its highest value at the center of the four quadrant parts of the lens. Figure 6(b) shows the distribution of intrinsic efficiency in one-dimensional direction for tapered channel lenses and square channel lenses. Compared with the square channel, the equivalent channel width of the tapered channel is smaller, so the acceptance range of the grazing incidence angle is smaller. Both types of channels reach the maximum intrinsic efficiency at 0.015 radian, and the intrinsic efficiency of the tapered channel is slightly higher than that of the square channel. Figure 7(a) shows the two-dimensional distribution of the intrinsic efficiency of the tapered multi-channel lens when the incident energy is 6.40 keV. At this time, the critical angle of total reflection of Ir without roughness is 0.013 radians, resulting in a truncation of the intrinsic efficiency of the lens within the original field of view, and the intrinsic efficiency of the channel at the truncation position reaches the maximum value. In Figure 7(b), the cut-off position of the two types of channel lenses is the same, and the maximum intrinsic efficiency of the tapered channel is slightly higher than that of the square channel.

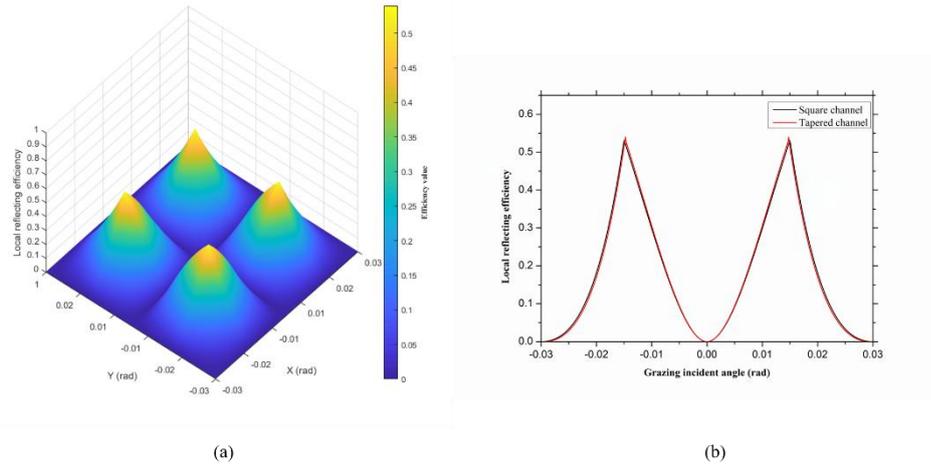

(a)                                                     (b)

Figure 6. (a) Two-dimensional distribution of intrinsic reflection efficiency (1.00keV) of the lens with tapered channel and curved surface; (b) One-dimensional distribution of intrinsic reflection efficiency (1.00keV) of two channel types of lenses

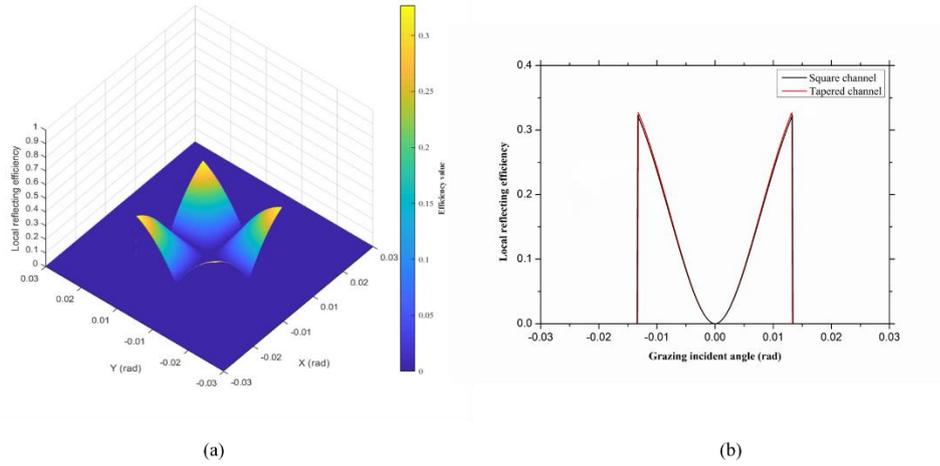

(a)                                                     (b)

Figure 7. (a) Two-dimensional distribution of intrinsic reflection efficiency (6.40keV) of the lens with tapered channel and curved surface; (b) One-dimensional distribution of intrinsic reflection efficiency of two channel type lenses (6.40keV)

### C. Spherical Aberration

The curved square multi-channel lens is similar to the flat type, and has spherical aberration when concentrating X-rays. The existence of spherical aberration makes X-rays not fully focused, and the focal spot has a certain width.

For a curved lens with a square channel, the exit width of the light beam is the same as the reflection width at the exit port of the multi-channel. In one dimension, at the ideal focal plane, the projected width is:

$$b(\theta) = \frac{\delta_{n\_sc}(\theta)\cos(\theta\chi_{tc})}{\cos(\theta\chi_{tc}+\phi)} \qquad (4\text{-}3)$$

The maximum value of the projection width is the width of the central focal spot, which is:

$$b_{max} = \frac{\cos(\theta\chi_{tc})}{\cos(\theta\chi_{tc}+\phi)} \begin{cases} t_{sc}|\theta_{max}|\chi_{sc}-(n-1)d_{sc}, & \frac{(n-1)}{\chi_{sc}t}d_{sc} < |\theta_{max}| < \frac{n}{\chi_{sc}t}d_{sc} \\ d_{sc}, & \frac{n}{\chi_{sc}t}d_{sc} < |\theta_{max}| < \frac{(n+1)}{\chi_{sc}t}d_{sc} \end{cases} \quad (4\text{-}4)$$

When the odd number of reflections is greater than or equal to 1, the maximum projected width is about $d_{sc}$. The angle between the square multi-channel satisfying the maximum projection width and the Z axis is $\phi_{sc} = nd_{sc}/\chi_{sc}t_{tc}-\theta_{max}$. At this time, the on-axis spherical aberration of the lens, that is, the depth of focus, is:

$$\Delta f = \frac{d_{sc}}{\tan(\theta_{max}+2\phi_{sc})} \quad (4\text{-}5)$$

In particular, consider the case where the parallel light is reflected once in each of the one-dimensional directions. The angle between the multi-channel satisfying the maximum projection width and the Z axis is $\phi_{sc} = d_{sc}/t_{sc}$, the vertical axis spherical aberration of the lens is $d_{sc}$, and the focal depth of the lens is $\Delta f = t_{sc}/2$.

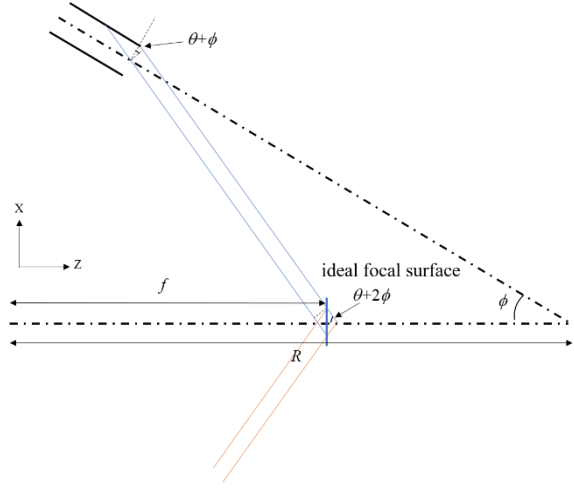

Figure 8. Schematic diagram of odd-numbered reflection projection of a square channel in one-dimensional direction

For curved lenses with tapered multi-channels, the situation is more complicated due to the different entrance and exit widths. The beam width at the exit is only related to the first angle of grazing incidence and the last angle of reflection, namely:

$$\delta_{n\_tc\_out}(\theta) = \frac{\delta_{n\_tc}(\theta)\cos(\theta\chi_{tc})}{\cos(\theta\chi_{tc}+2\alpha_{tc})} \quad (4\text{-}6)$$

Figure 9 is a schematic diagram of X-rays after an odd number of reflections in a conical multi-channel and finally converged on an ideal focal plane. When the X-ray is reflected only once in the channel, the maximum value of the projection width is:

$$b_{max} = \frac{\cos^2(\theta_{max}\chi_{tc})}{\cos(\theta_{max}\chi_{tc}+\phi)\cos(\theta_{max}\chi_{tc}+2\alpha_{tc})} \begin{cases} t_{tc}\tan(\theta_{max}\chi_{tc}), & 0<|\theta_{max}|<\dfrac{d_b}{\chi_{tc}t_{tc}} \\ d_b, & \dfrac{d_b}{\chi_{tc}t}<|\theta_{max}|<\dfrac{d_b+d_f}{\chi_{tc}t_{tc}} \end{cases}$$

When the conditions are met:

$$\begin{cases} \theta_1\chi_{tc}=\dfrac{d_b}{t_{tc}} \\ \theta_2\chi_{tc}=-\dfrac{d_f}{t_{tc}} \end{cases} \quad (4\text{-}7)$$

The projection width of the first reflected beam on the focal plane reaches the maximum value, which is $d_b$. Correspondingly, from the geometric relationship, the focal depth of light convergence is:

$$\Delta f = \frac{d_b}{2\tan(\theta_{max1}+2\phi_{sc1})} + \frac{d_b}{2\tan(|\theta_{max2}+2\phi_{sc2}|)} \quad (4\text{-}8)$$

In particular, consider the case where the parallel light is reflected once in each of the one-dimensional directions. At this point, $\theta_1=|\theta_2|=0$ is satisfied. The angles between the multi-channel satisfying the maximum projection width and the Z axis are $\phi_{tc1}=d_b/t_{tc}$ and $|\phi_{tc2}|=d_f/t_{tc}$, respectively, the vertical axis spherical aberration of the lens is $d_b$, and the focal depth of the lens is $\Delta f = \tfrac{1}{4}t_{tc}(1+d_b/d_f)$. It can be seen that, when the parallel light source is focused without considering the wall thickness, the vertical and axial spherical aberrations formed by the two-channel lenses are both cone-shaped after each reflection in the two-dimensional direction. The channel is smaller than the square channel.

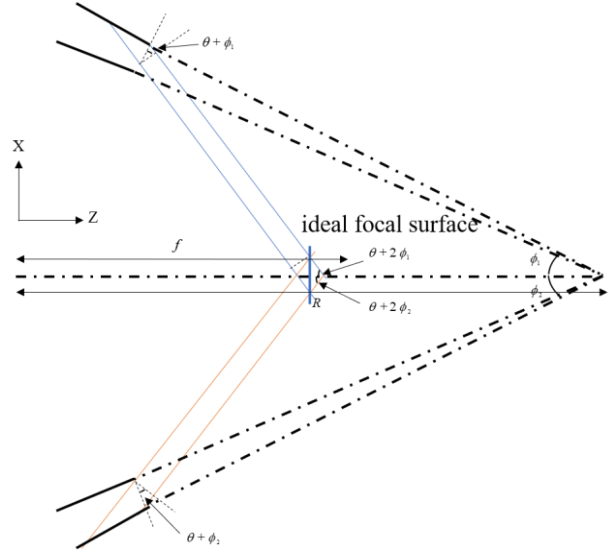

Figure 9. Schematic diagram of odd-numbered reflection projection of one-dimensional conical channel

Consider again the case where the light is reflected multiple times ($n \geq 3$) within the cone-shaped channel. Since the maximum value of the reflection width associated with the upper wall is greater than the reflection width associated with the lower wall, the calculation method of the spherical aberration formed by the multiple reflections of the tapered multi-channel is similar to that of the square channel. The maximum value of the projection width is:

$$b_{\max} = A_{tc} \begin{cases} t_{tc}^{(n-2)'}(\theta_{\max}) \tan(\theta_{\max} \chi_{tc}) - d_b^{(n-2)'}(\theta_{\max}), \\ \qquad \dfrac{(d_f + d_b^{(n-3)'}(\theta_{tcm|n-3}))}{\chi_{tc} t_{tc}^{(n-3)'}(\theta_{tcm|n-3})} < |\theta| < \dfrac{(d_f + d_b^{(n-2)'}(\theta_{tcm|n-2}))}{\chi_{tc} t_{tc}^{(n-2)'}(\theta_{tcm|n-2})} \\ t_{tc}^{(n-2)'}(\theta_{\max}) \dfrac{(d_f + d_b^{(n-2)'}(\theta_{tcm|n-2}))}{t_{tc}^{(n-2)'}(\theta_{tcm|n-2})} - d_b^{(n-2)'}(\theta_{\max}), \\ \qquad \dfrac{(d_f + d_b^{(n-2)'}(\theta_{tcm|n-2}))}{\chi_{tc} t_{tc}^{(n-2)'}(\theta_{tcm|n-2})} < |\theta| < \dfrac{(d_f + d_b^{(n-1)'}(\theta_{tcm|n-1}))}{\chi_{tc} t_{tc}^{(n-1)'}(\theta_{tcm|n-1})} \end{cases} \quad (4\text{-}9)$$

$A_{tc}$ is slightly less than 1.

For multiple reflections, the maximum projection width is $p_{\max}$, then the depth of focus is:

$$\Delta f = \frac{p_{\max}}{\tan(\theta_{\max 1} + 2\phi_{sc1})} \quad (4\text{-}10)$$

When the X-ray is reflected too many times in one-dimensional direction, the incident angle of the $n+1$ reflection satisfies the condition $\theta^{n+1}\chi_{tc} \geq \dfrac{\pi}{2}$, and the light begins to deflect the propagation direction and propagate in the negative direction of the axis. At this time, the light cannot be focused on the focal plane, and the field of view of the lens is limited. At this time, the spherical aberration of the lens is determined by the low-order reflection.

**D. Point-spread Function(PSF)**

The PSF of a lens describes its systematic response to a point source and is a measure of the imaging performance of the lens.

Given a point isotropic point light source, it is assumed that it emits N photons per second within the solid angle of $4\pi$, and the light beam emitted from the light source at the emission angle $(\theta_x, \theta_y)$ has $(n_x, n_y)$ total reflections in a certain channel of the curved lens, which is in the focus. The light intensity distribution on the surface is:

$$i(\theta_x, \theta_y) = \dfrac{N}{4\pi(1+M_T)^2 l_s^2} R^{nx}(\theta_x \chi) R^{ny}(\theta_y \chi) \tag{4-11}$$

For a curved lens with a square multi-channel, the maximum emission angle of light and the maximum number of reflections satisfy:

$$\dfrac{(m-1)d}{\chi_{sc} t} < \theta_{max} < \dfrac{(m+1)d}{\chi_{sc} t} \tag{4-12}$$

The structure factor and coordinate factor of the lens are taken as:

$$\begin{cases} \alpha = t_{sc} \theta_{max} / d_{sc} \\ z = 2|x| / d_{sc} \end{cases} \tag{4-13}$$

When the number of reflections is $i = m$, the emission angle interval is:

$$\begin{cases} X_m(x) = \dfrac{\theta_{max}}{\chi_{sc}} \left[ \min\left(1, \dfrac{m-1+z}{\alpha}\right), 1 \right], & m-1 < \alpha < m \\ X_m(x) = \dfrac{\theta_{max}}{\chi_{sc}} \left[ \dfrac{m-1+z}{\alpha}, \min\left(1, \dfrac{m+1-z}{\alpha}\right) \right], & m < \alpha < m+1 \end{cases} \tag{4-14}$$

When the number of reflections is $i \leq m-2$, the emission angle interval is:

$$X_i(x) = \dfrac{\theta_{max}}{\chi_{tc}} \left[ \min\left(1, \dfrac{i-1+z}{\alpha}\right), \min\left(1, \dfrac{i+1-z}{\alpha}\right) \right] \tag{4-15}$$

For the direct case, the emission angle interval is:

$$X_0(x) = \begin{cases} \dfrac{\theta_{\max}}{\chi_{sc}}\left[1, \dfrac{1-z}{\alpha}\right], & 0 < \alpha < 1 \\ \dfrac{\theta_{\max}}{\chi_{sc}}\left(0, \dfrac{1-z}{\alpha}\right), & \alpha > 1 \end{cases} \quad (4\text{-}16)$$

The same as the flat-panel lens, set the central focal point of the focal plane, the cross focal line and the background X-rays are OO, OE (EO) and EE X-rays, respectively, and their intensity distribution is the reflection of incident X-rays in the emission angle range in two-dimensional directions The product of rate integrals, that is:

$$\begin{aligned}
I_{oo}(x,y) &= \dfrac{N\eta\theta_{\max}^2}{\pi(1+M_T)^2 d_{sc}^2 \chi_{sc}^2} f_o(z_x;\alpha) f_o(z_y;\alpha), \\
I_{oe}(x,y) &= \dfrac{N\eta\theta_{\max}}{4\pi l_s^2 (1+M_T)^2} R^{n_y}\left(\dfrac{y}{(1+M_T)l_s\chi_{sc}}\right) \times \dfrac{2l_s}{d_{sc}} f_o\left(\dfrac{2|x|}{d_{sc}};\alpha\right), \\
I_{eo}(x,y) &= \dfrac{N\eta\theta_{\max}}{4\pi l_s^2 (1+M_T)^2} R^{n_x}\left(\dfrac{x}{(1+M_T)l_s\chi_{sc}}\right) \times \dfrac{2l_s}{d_{sc}} f_o\left(\dfrac{2|y|}{d_{sc}};\alpha\right), \\
I_{ee}(x,y) &= \dfrac{N\eta}{4\pi l_s^2 (1+M_T)^2} R^{n_x}\left(\dfrac{x}{(1+M_T)l_s\chi_{sc}}\right) R^{n_y}\left(\dfrac{y}{(1+M_T)l_s\chi_{sc}}\right)
\end{aligned} \quad (4\text{-}17)$$

For curved lenses with tapered multi-channels. Take the average value of the entrance width and exit width as the equivalent width of the lens: $d_{tc} = (d_f + d_b)/2$, the structure factor and coordinate factor of the lens are:

$$\begin{cases} \alpha = t_{tc}\theta_{\max}/d_{tc} \\ z = 2|x|/d_{tc} \end{cases} \quad (4\text{-}18)$$

When $m \leq 2$, the intensity distribution of various types of light can be obtained.

Consider again the situation of $m > 2$. From the equation $\delta_{n\_tc}(\theta) = x$, the upper and lower bounds of the emission angle interval can be obtained when $i = m$, under the constraints. Since the upper and lower bounds of the emission angle interval are related to the structure factor, the coordinate factor and the maximum number of reflections, they are denoted as $f_{r\_u}(z,\alpha;m)$ and $f_{r\_d}(z,\alpha;m)$, respectively. Depend on:

$$\dfrac{(d_f + d_b^{(m-2)'}(\theta_{tcm|m-2}))}{\chi_{tc} t_{tc}^{(m-2)'}(\theta_{tcm|m-2})} = \alpha \quad (4\text{-}19)$$

It can be solved that the structure factor judgment value M is used to judge the last reflection and the interval range of the emission angle:

$$\begin{cases} X_m(x) = \dfrac{\theta_{max}}{\chi_{tc}}\left[\min\left(1, f_{r\_d}(z,\alpha;m)\right), 1\right], & M-1 < \alpha < M \\ X_m(x) = \dfrac{\theta_{max}}{\chi_{tc}}\left[f_{r\_u}(z,\alpha;m), \min\left(1, f_{r\_u}(z,\alpha;m)\right)\right], & M < \alpha < M+1 \end{cases} \quad (4\text{-}20)$$

When $1 < i \leq m\text{-}2$, the emission angle interval is:

$$X_m(x) = \dfrac{\theta_{max}}{\chi_{tc}}\left[\min\left(1, f_{r\_d}(z,\alpha;m)\right), \min\left(1, f_{r\_u}(z,\alpha;m)\right)\right] \quad (4\text{-}21)$$

For the direct case, the emission angle interval is:

$$X_0(x) = \begin{cases} [0, \theta_{max}], & 0 < \theta_{max} < M_0 \\ \left[0, \dfrac{d_f - x}{t_{tc}\chi_{tc}}\right], & \theta_{max} > M_0 \end{cases} \quad (4\text{-}22)$$

Integrate the odd and even reflectivities in the emission angle interval for $m$ reflections to obtain $f_o(z;\alpha)$ and $f_e(z;\alpha)$, respectively.

**E. Collection Efficiency of Incident X-ray**

Figure 10 shows the relationship between the collection efficiency and the structure factor $\alpha_{sc}$ of the curved lens of the square multi-channel for various types of light. With the increase of the structure factor value, the collection efficiency of the curved lens for OO light first increases and then fluctuates and approaches 0.2.

Figure 11 is a graph showing the relationship between the collection efficiency and the structure factor $\alpha_{tc}$ of the curved lens with the tapered multi-channel for various types of light. When $\alpha = \sqrt{2}$, the maximum collection efficiency of the lens for OO light is the highest. Overall, the collection efficiency curves for each type of light are similar to square channels. In particular, with the increase of the structure factor, the collection efficiency of OO light and OE or EO light by the tapered channel decreases faster than that of the square channel. This means that when the geometrical parameters of the two channel shapes are the same, ie $d_{sc} = d_{tc}$, $t_{sc} = t_{tc}$, with the increase of the incident X-ray energy, the square channel has a higher X-ray collection efficiency and a larger effective field of view.

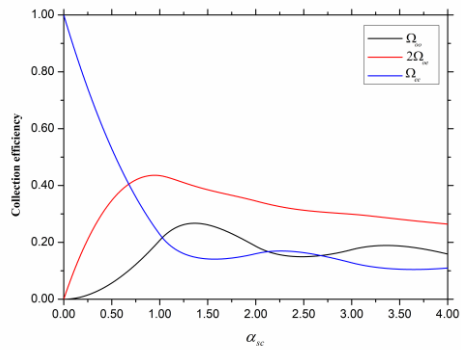

Figure 10. The relationship between the light collection efficiency and the structure factor of the square multi-channel curved lens

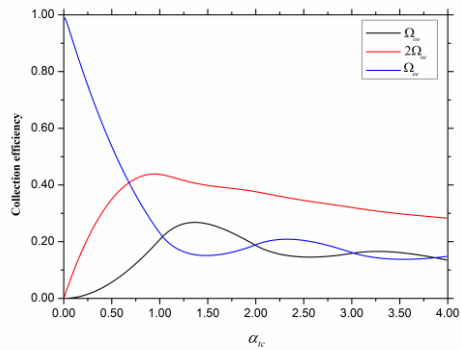

Figure 11. The relationship between the light collection efficiency and the structure factor of the tapered multi-channel curved lens

Figure 12(a) shows the two-dimensional intensity distribution of the central focal spot calculated by simulation when the lens focuses a parallel light source with an energy of 6.4 keV. When the incident energy is 6.4keV, the X-ray can only undergo one total reflection in one-dimensional direction, and the distribution of OO X-rays is "triangular", as shown in Figure 12(b). The full width at half maximum of the intensity curve is about 15 $\mu m$.

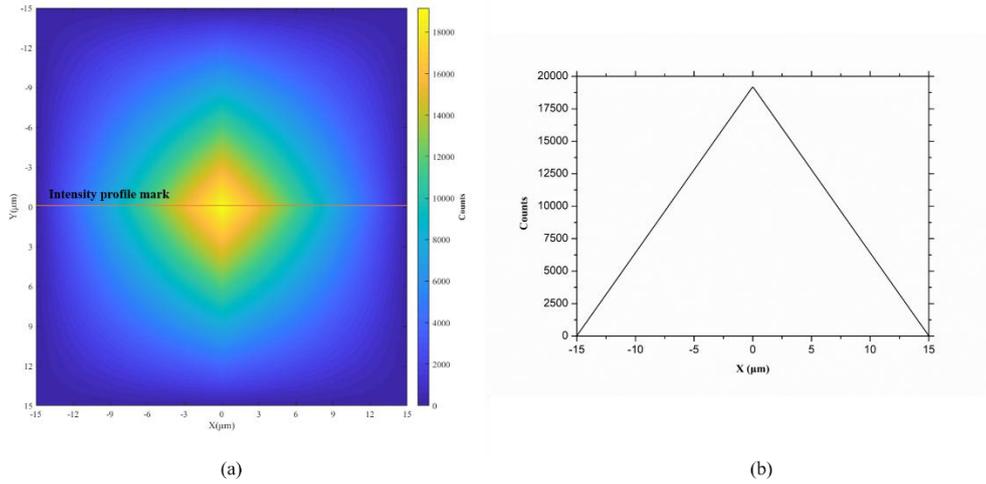

(a)                                              (b)

Figure 12. Light intensity distribution (6.4keV) of OO light in a square multi-channel curved lens: (a) two-dimensional distribution; (b) one-dimensional distribution

Figure 13(a) and Figure 13(b) are respectively the two-dimensional distribution of the central focal spot and the intensity distribution in one-dimensional direction of the 6.4keV parallel light source focused by the curved lens of the conical multi-channel. The full width at half maximum of the central focal spot intensity curve of the tapered multi-channel is about 14.9 $\mu m$, which is slightly narrower than that of the square multi-channel.

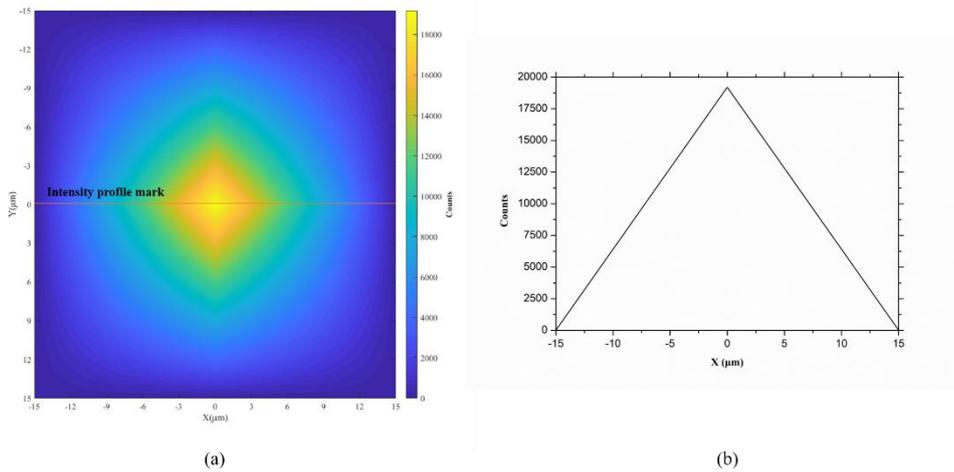

(a)                                              (b)

Figure 13. Light intensity distribution (6.4keV) of OO light in the conical multi-channel curved lens: (a) two-dimensional distribution; (b) one-dimensional distribution

When the curvature factor is determined, the collection efficiency of the OO light by the curved lens is only related to the structure factor. Figure 14 shows the two-dimensional distribution of OO light with incident energy and aspect ratio. The interface material is Ir, the channel shape is square, and the curvature factor is 3. Also, affected by the absorption edge, the collection efficiency fluctuates significantly near the absorption edge. In the preliminary basic design of the surface, the channel aspect ratio can be selected according to the incident energy.

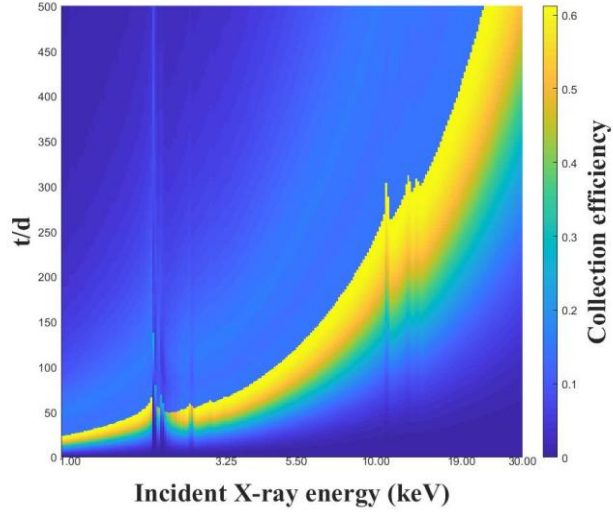

Figure 14. Two-dimensional distribution of OO light collection efficiency of curved lens

**F. Peak-to-background Ratio on the Focal Plane**

In the flat panel, with the increase of the incident X-ray energy, the focal peak-to-background ratio of the cross focal plane has the greatest influence. Therefore, when evaluating the focusing efficiency of the focal plane of the curved lens, the ratio of the average intensity of OO light to the average intensity of OE light is used to calculate, then the formula (4-74) becomes:

$$\frac{\overline{I_{oo}}}{\overline{I_{oe}}} = \begin{cases} \dfrac{2(1+M_T)l_s\theta_{max}\eta\Omega_{oo}}{\chi b_{max}\Omega_{oe}}, & Point\ Light \\ \dfrac{\eta\Omega_{oo}}{b_{max}\Omega_{oe}}, & Directional\ Light \end{cases} \quad (4\text{-}23)$$

It can be seen from the above formula that the peak back of the focal plane is inversely proportional to the curvature factor. For a lens with a given radius of curvature, in order to enable it to have a higher peak-to-background ratio of the focal plane, it is mainly necessary to consider the effects of incident energy and aspect ratio. In addition, since the interface reflectivity uses a two-factor model, the effect of roughness also needs to be considered. The following is an example of focusing a point light source with a lens with a square multi-channel and a radius of curvature of 1500mm, where $l_s = R/2$.

Figure 15 shows the curve of the focal plane peak-to-back ratio at different energies as a function of the structure factor when the interface material is Ir and there is no roughness. With the increase of the incident X-ray energy, when the value of the structure factor is constant, the peak-to-background ratio decreases. When the curvature and aspect ratio of the curved lens are the same, the higher the incident X-ray energy is, the more the focusing performance of the curved lens is affected by the cross focal line, and the lower the imaging surface resolution.

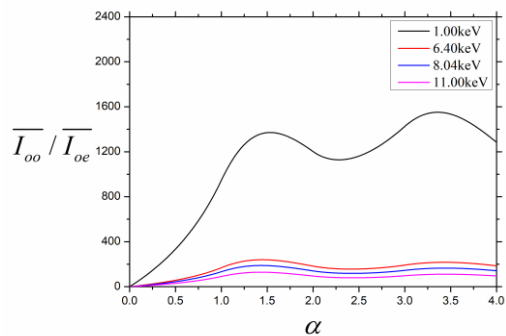

Figure 15. Relationship between peak-to-background ratio and structure factor of curved lens under different incident energies

Figure 16 shows the curve of the peak-to-back ratio of the focal plane as a function of the structure factor under different roughness conditions when the interface material is Ir and the incident energy is 6.4 keV. With the increase of surface roughness, the peak-to-back ratio decreases, the cross focal line affects the focusing performance of the lens, and the imaging resolution decreases under the same structure factor value.

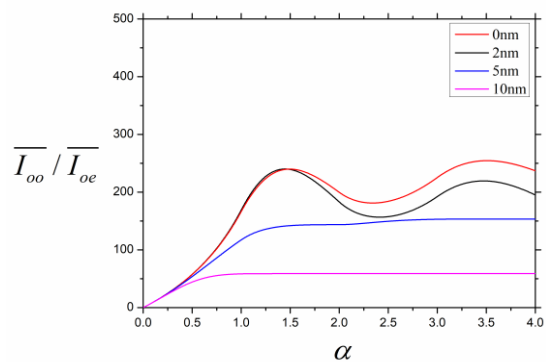

Figure 16. The relationship between the peak-to-background ratio and the structure factor of the curved lens with the incident energy of 6.4keV and different roughness

## 5. Conclusion

The structural principles and simulated X-ray paths of both the FMCP and the CMCP were described by the MATLAB, By the Chapman Model, the field of view, local reflection efficiency, spherical aberration, point-spread function, collection efficiency of incident X-ray and peak-to-background ratio on the focal plane of the two devices were compared. Therefore, it can be concluded that the FMCP cannot focus parallel X-ray but the collection efficiency was higher then the CMCP. The CMCP can focus the parallel X-ray, and the local reflection efficiency was higher then the FMCP but the imaging resolution was lower then it. In the future, these devices could be selected according to the different field needs.

**Acknowledgment**
This work was supported by the National Natural Science Foundation of China (Grant No. 12175021), the National Natural Science Foundation of China (Grant No. 12175254) and the Guangdong Basic and Applied Basic Research Foundation (Grant No. 2019A1515110398).